\documentclass[pre,twocolumn,aps,floatfix]{revtex4}
\usepackage{times}
\usepackage{bm}
\usepackage{amsmath,amssymb}
\usepackage{graphicx}
\begin{document}
\date{\today}
\title{Colloidal charge reversal: dependence on the ionic size and the electrolyte concentration}
\author{Alexandre Diehl} 
\email{diehl@ufpel.edu.br}
\affiliation{Departamento de F\'{\i}sica, Instituto de F\'{\i}sica e Matem\'atica,
Universidade Federal de Pelotas, Caixa Postal 354, CEP 96010-900, Pelotas, RS, Brazil}
\author{Yan Levin} 
\email{levin@if.ufrgs.br}
\affiliation{Instituto de F\'{\i}sica, Universidade Federal
do Rio Grande do Sul\\ Caixa Postal 15051, CEP 91501-970, 
Porto Alegre, RS, Brazil}
\date{\today}

\begin{abstract}
Extensive Monte Carlo simulations and scaling arguments are used to study 
the colloidal charge reversal. The critical colloidal surface charge density $\sigma_c$ at
which the reversal first appears is found to depend strongly on the ionic
size.  We find that  $\sigma_c$ has an inflection point as a function of the 
electrolyte concentration.  The width of the plateau region in the vicinity of the inflection
point depends on the temperature and the ionic radius $a$. 
In agreement with the theoretical predictions
it is found that the critical colloidal charge 
above which the electrophoretic mobility 
becomes reversed diverges as $Z_c \sim 1/a^2$ in the limit  $a \rightarrow 0$.   
\end{abstract}

\maketitle

\section{Introduction}
\label{introd}

A common way to stabilize colloidal suspensions against flocculation and
precipitation
is by synthesizing particles with acidic groups on 
their surface.  When placed in water, these groups dissociate and colloids
acquire a net negative charge.  In aqueous suspension containing only monovalent
electrolyte, the long range Coulomb repulsion prevents the colloids
from approaching one another to distances for which the short range
van der Waals interaction can lead to an irreversible sticking and precipitation.
However, when besides the 1:1 electrolyte, suspensions contain some multivalent
counterions, a number of curious and very counterintuitive effects can take 
place~\cite{Le02}.  For example,
it has been observed that in such suspensions two like-charged colloidal particles
can attract one 
another~\cite{Pa80,KjMa86,GuJoWe84,CrGr94,GrMaBr97,AlAmLo98,LaPiLeFe00,Gr00,DiCaLe01,LoLyLi01,NaNe04,KoLe99,RoBl96,GeBrPi00,KoLe07}.  
This attraction is not a result of the van der 
Waals interaction,  but rather a consequence of strong positional correlations between
the multivalent counterions surrounding the colloidal 
particles~\cite{StRo90,DiTa99,Le99a,ArStLe99,SoCr99}.  The mechanism
of this attraction has been studied extensively, yet a fully predictive theory of this
phenomenon still remains elusive.  

Another curious effect observed in dilute colloidal
suspensions containing multivalent counterions is the reversal 
of the electrophoretic mobility~\cite{TaGr01,GrNgSh02,Le02,So02}.  
Since the bare charge of colloidal particles is negative,
when the electrostatic potential gradient is established in the suspension, one naturally expects
that the particles should move in the direction {\it opposite} to the established electric field.
Yet, what is often found is quiet the 
opposite  --- particles drift in the direction of the field~\cite{QuCaHi02,MaQuGa03,FeFe05}.
The reversal of the electrophoretic mobility is a consequence of strong 
electrostatic interaction between the colloidal particles and the multivalent 
counterions~\cite{ToVa80,LoSaHe82,DeJi01}.  As a
consequence of this coupling, some counterions become associated (condensed) 
with the colloidal particle.  The  positional correlations induced by 
the electrostatic repulsion between the condensed counterions 
can lead to colloid-counterion complexes which are overcharged (charge-reversed)  --- the
number of condensed counterions can actually be larger than is necessary to completely 
neutralize the colloidal charge~\cite{Sh99a,SoCr01a,MeHoKr01,Le02,PiBaLe05,LeHo08}.  
If this happens, the electrophoretic mobility of colloidal particles will be reversed. 
While there are some theories which qualitatively account for this curious behavior, 
no fully predictive approach is yet available~\cite{PiBaLe05}.  
In this paper we will use extensive Monte Carlo simulations
to explore two aspects of this problem --- 
the dependence of the minimum colloidal charge at which the reversal
of the electrophoretic mobility takes place on: (one) the 
concentration of the multivalent $z$:1 electrolyte
and (two) on the ionic size.

\section{The model and simulations}
\label{model}

The  electrophoretic mobility, in general, is a complicated 
non-linear function of the electrokinetic $\zeta$ potential~\cite{RuSaSc89,JoYbTr04}.
For small $\zeta$ and large ionic strengths, however, 
the relationship between the two is linear and is given by the
Smoluchowski equation~\cite{Hu81,RuSaSc89}. A change in the
sign of the $\zeta$ potential will, therefore, lead to the reversal of
the electrophoretic mobility.  which we will also associate 
with the overcharging (or the charge-reversal) of the colloidal particles. In principle,
the overcharging (charge-reversal) and the reversal of the electrophoretic mobility
are two distinct concepts -- one is static and the other dynamical.  In practice, however,
the definition of charge reversal carries some ambiguity.  
The general trends, such as the
behavior of the effective charge and of the electrophoretic mobility as a 
function of, say, the
ionic size or electrolyte concentration are very similar in two cases~\cite{DiLe06}. We will,
therefore, use the two concepts interchangibly.   
For a fixed electrolyte concentration, the
value of the colloidal charge at which the $\zeta$-potential vanishes will be designated
as the {\it critical colloidal charge}.  Our goal is to find the dependence of this charge
on the ionic size and the electrolyte concentration.
 
We consider a diluted aqueous mixture of colloidal particles 
inside a $z$:1 electrolyte. The spherical colloidal particles have radius $R$
surface charge $-Zq$, where $q$ is the elementary charge.  
For each colloidal particle there are
$Z$ monovalent counterions. 
All the ions are modeled as hard spheres 
of radius $a$ with the charge $+zq$ (electrolyte counterion), $-q$ (electrolyte coion) or 
$+q$ (colloidal counterion) at their centers. 
The solvent is treated as a continuum of dielectric constant $\varepsilon$. 
The relative strength of the electrostatic interactions, as compared to the
thermal energy, is measured by the ratio of the ionic radius to the 
Bjerrum length, $\lambda_B = q^2/4\pi \varepsilon k_BT$.

As was argued in Ref. ~\cite{DiLe06}, 
the value of the $\zeta$ potential may
be associated with the electrostatic potential at the effective
shear plane removed from the colloidal surface by one ionic diameter~\cite{LoDu06}. Since the
maximum of the electrostatic potential also occurs at approximately the same position, the
precise location of the shear plane does not influence strongly the 
value of the $\zeta$-potential.  In this respect, our approach is quite similar to the one 
adopted by Bjerrum for simple electrolytes~\cite{Bj26,FiLe96}.  
The static potential at the effective shear
plane can then  be calculated using the canonical Monte Carlo (MC) 
simulations~\cite{DiLe06}. Recent simulations show the basic correctness of this picture for
normal wetting surfaces~\cite{JoYbTr04,DiLe06}.  Working with $\zeta$-potential is also 
advantageous as compared to defining the effective charge in terms of
condensed counterions located within a sheath surrounding the  
colloidal surface.  Such definition carries a large degree of arbitrariness,
since the condensed counterions will in turn drive a co-associations of coions.  The effective
charge will then be strongly sensitive to the precise value of the sheath width.  This is
not the case for $\zeta$-potential, which under the same conditions develops a maximum near
the colloidal surface which diminishes its sensitivity to the 
precise location of the shear plane.  Furthermore, since the $\zeta$-potential is calculated
by integrating the electric field over the whole space, it already takes into account the
layering effect that hinders the geometrical definition of the effective charge.

A colloidal particle is fixed at 
the center of a cubic simulation box of side length $L$ 
and is surrounded by the 
counterions and coions, 
the number of which satisfies the overall charge neutrality. We define 
$C$ as the molar concentration of the $z$-valent counterions derived from the dissociation of 
$z$:1 (strong) electrolyte --- assumed to be fully dissociated in an aqueous environment.  
The electrostatic interactions 
are computed using the Ewald summation method~\cite{AlTi87} with 518 Fourier-space wave vectors and 
a real-space damping parameter $\kappa = 5/L$. 

Two types of MC moves were utilized --- ion transfer to a completely new random position 
inside the simulation box, which is useful for low salt concentrations, and a small 
linear displacement for high salt concentration, in order to give the standard 
acceptance ratios for the Metropolis algorithm. The number of microions in 
each simulation was varied from approximately 50 up to 3000 particles, depending on 
the molar salt concentration and the box length. Typical runs involved 10$^7$ Monte Carlo 
steps for equilibration and 10$^8$ steps for production. 
After equilibration, the average number of 
counterions and coions in concentric spherical shells of equal thickness around 
the colloid were accumulated in order to obtain the density 
profiles $\rho_i(r)$.
The mean electrostatic potential at distance $r$ from the colloidal particle is 
then calculated as
\begin{equation}
\label{e1}
\phi (r) = \int_r^{\infty} {\rm d}r'\; E(r')=
\frac{q}{4\pi \varepsilon}\, \int_r^{\infty} {\rm d}r'\; \frac{P(r')}{r'^2}\;,
\end{equation}
where $E(r)$ is the electric field and $P(r)$ is the integrated charge (in units of $q$)
within a distance $r$ from the center of the colloidal particle, 
\begin{equation}
\label{e2}
P(r) = -Z + \int_{R}^r \left[ \sum_i z_i \rho_i(r')\right]4\pi r'^2dr'\;,
\end{equation}
where $i$ refers to the type of the microion.
Since the typical integrated charge rapidly decays to zero~\cite{DiLe06}, the 
upper cutoff in Eq.~(\ref{e1}) is taken to be $L/2$.
Following the Ref. ~\cite{DiLe06}, the shear plane was located 
at one ionic diameter from the colloidal surface, so that 
$\zeta \equiv \phi(R_s)$, where $R_s = R + 2a$. 

\section{Scaling analysis}

The model presented in the previous section is quite complex, with a number of distinct
length scales.  To organize and interpret the data of the Monte Carlo simulations 
we shall, therefore, appeal to the dimensional and scaling analysis. There are five basic
length scales: $R$, $L$, $a$, $\lambda_B$, and $C^{-1/3}$.  Since we are interested in very dilute
suspensions, $L \rightarrow \infty$.  Although this limit can not be achieved in the simulations,
our box size was always taken to be sufficiently large so that critical colloidal charge 
$-Z_c q$  did not have 
any explicit dependence
on $L$. We are, therefore, left with four relevant length scale, so that 
for a fixed $z:1$ electrolyte,
$Z_c$ is a function of only three dimensionless ratios
\begin{equation}
\label{e3}
Z_c =f\left(\frac{a}{R}, \frac{a}{\lambda_B}, \lambda_B C^{1/3} \right)\;,
\end{equation}
where $f(x,y,z)$ is a scaling function. 
Furthermore, we note that when the 
Debye length $\xi_D=1/\sqrt{4 \pi \lambda_B (z^2+z) C}$, is sufficiently short,
$R/\xi_D \gg 1$  --- which is almost always the case near the isoelectric point --- 
the curvature effects will be screened, 
and the critical colloidal charge must be  
proportional to the colloidal surface area.  This means that
\begin{equation}
\label{e4}
Z_c =\frac{4 \pi R^2}{a^2} g\left(\frac{a}{\lambda_B}, \lambda_B C^{1/3} \right)\;.
\end{equation}
We conclude that for sufficiently large salt concentrations, the reversal of the electrophoretic
mobility will take place when the modulus of the 
colloidal surface charge density, $\sigma \equiv Zq/4 \pi R^2$,
is larger than the critical value $\sigma_c$, which depends on the Bjerrum length, ionic radius, and 
the concentration of electrolyte through the scaling function $g(x,y)$,
\begin{equation}
\label{e5}
\sigma_c=\frac{q}{a^2} g\left(\frac{a}{\lambda_B}, \lambda_B C^{1/3} \right)\;.
\end{equation}
The similarity transformation, Eq.~(\ref{e5}), is particularly useful when one wants to obtain
the critical surface charge density for suspensions with large concentrations of electrolyte. 
In these cases, the direct MC simulations become extremely slow, due to large number
of microions which must be used to simulate a {\it dilute} colloidal suspension in the 
$L \rightarrow \infty$ limit.  However,  Eq.~(\ref{e5}) 
tells us that this critical surface charge density
can also be obtained by simulating a much smaller system at a slightly lower temperature and with
a somewhat larger microions.  For example, suppose that we want to find the critical surface
charge density of colloidal particles inside a dilute suspension at room 
temperature, $\lambda_B=7.2\,\mbox{\AA} \,$, containing 3:1 electrolyte
at concentration $C=1\,$M, with ions of radius $2\,\mbox{\AA} $.  
Instead of doing the direct simulation of this system, we can
simulate a ``similar''system with say half the number of microions, $C=0.5\,$M, at a slightly
lower temperature, $\lambda_B=7.2\times 2^{1/3}=9.07 \,\mbox{\AA}$ and with ions 
of radius $2\times 2^{1/3}= 2.5 \,\mbox{\AA}$.  From Eq.~(\ref{e5}), the critical
charge of the original system (at concentration $C=1\,$M) will be  $2^{2/3}$ times the critical charge
of the ``similar'' system. The latter simulations, however, are much easier to perform since
the number of microions involved is much smaller.

In Refs.~\cite{Le04} and ~\cite{PiBaLe05} 
it was argued that the critical surface charge density is determined by the
work that must be performed to transfer a multivalent counterion from the bulk electrolyte 
to the colloidal surface.  In particular, it was found 
that $\sigma_c \sim (\Delta \mu)^2$, where $\Delta \mu$ is the change in
the ionic solvation free energy between the bulk and the colloidal surface~\cite{Le04}.  
In the limit of vanishing $a$,  $\Delta \mu$ diverges as 
$1/a$.  This divergence is a consequence of Bjerrum pairing of oppositely
charged ions in the bulk electrolyte~\cite{Bj26,LeFi96,Le02}.  
The critical surface charge density, therefore, 
diverges as  $\sigma_c \sim 1/a^2$.  Thus, the 
scaling function with $x=0$, $g(0,y)$, 
should be constant for all values of $y$. For small, but finite values
of $x$, we expect to see a deviation from this behavior in two limits: when the entropic 
effects begin to dominate the electrostatics $C^{1/3} \lambda_B < 0.5$, 
and when the hard-core repulsion 
begins to dominates everything, $C^{1/3} a > 0.2$.

\section{Results and discussion}
\label{discussion}

\begin{figure}[t]
\includegraphics[width=8cm]{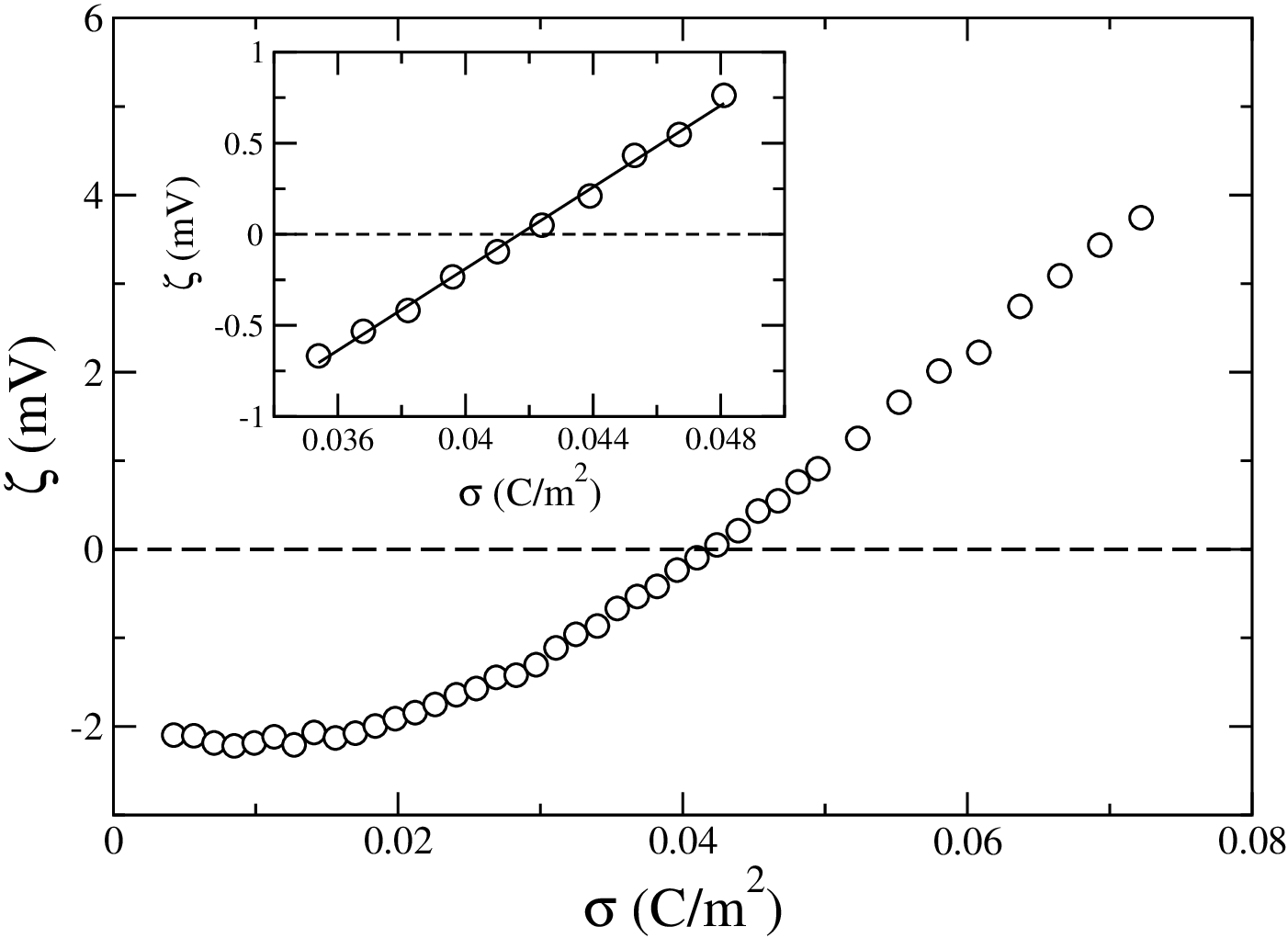}
\caption{Zeta potential as a function of the colloidal surface charge density. The molar 
concentration of 3:1 electrolyte is $C=0.1\,$M, the radius of the microions is 
$a=3\,$\AA, and the Bjerrum length is $7.2\,\mbox{\AA}$. The inset shows the region 
where the $\zeta$ potential becomes reversed, $\zeta = 0$, with a very good linear fit 
(solid line) to the simulation data (circles), from which the 
precise value of $\sigma_c$ is determined.}
\label{fig1}
\end{figure}

In Fig.~\ref{fig1} we show the $\zeta$-potential as a function of the colloidal charge 
density for a suspension containing 3:1 electrolyte at a molar concentration 
of $C=0.1\,$M. The radius of the colloidal particle was fixed at $R=30 \,\mbox{\AA}$, while 
its charge $Z$ was varied. 
Most of the simulations were performed with the boxes of side length $L=120\,\mbox{\AA}$, 
which is large enough to produce very small colloidal volume fractions, thus, minimizing 
the influence of the periodicity on the ionic distribution. However, 
for low salt concentrations we increased the box length to $L=150\,\mbox{\AA}$ and $210\,$\AA, 
in order to increase the number of particles and to obtain a  
better statistics of ion distribution around the colloid.

For weakly charged colloidal particles the increase 
(in modulus) of the surface charge density was accompanied by a uniform decline of the
$\zeta$ potential ($\zeta$ accompanied the colloidal charge and became more negative). 
However, when the colloidal charge became sufficiently large, 
counterion condensation became important 
and $\zeta$ {\it increased} as a function of the bare colloidal charge, 
becoming positive for sufficiently strongly charged colloids, see Fig.~\ref{fig1}. 
To accurately determine 
the critical  colloidal charge density $\sigma_c$ at which $\zeta=0$, 
we used a linear interpolation of the 
simulation data, as shown in the inset of Fig.~\ref{fig1}.\\    

\vspace*{0.5cm}
\begin{figure}[hb]
\includegraphics[width=8cm]{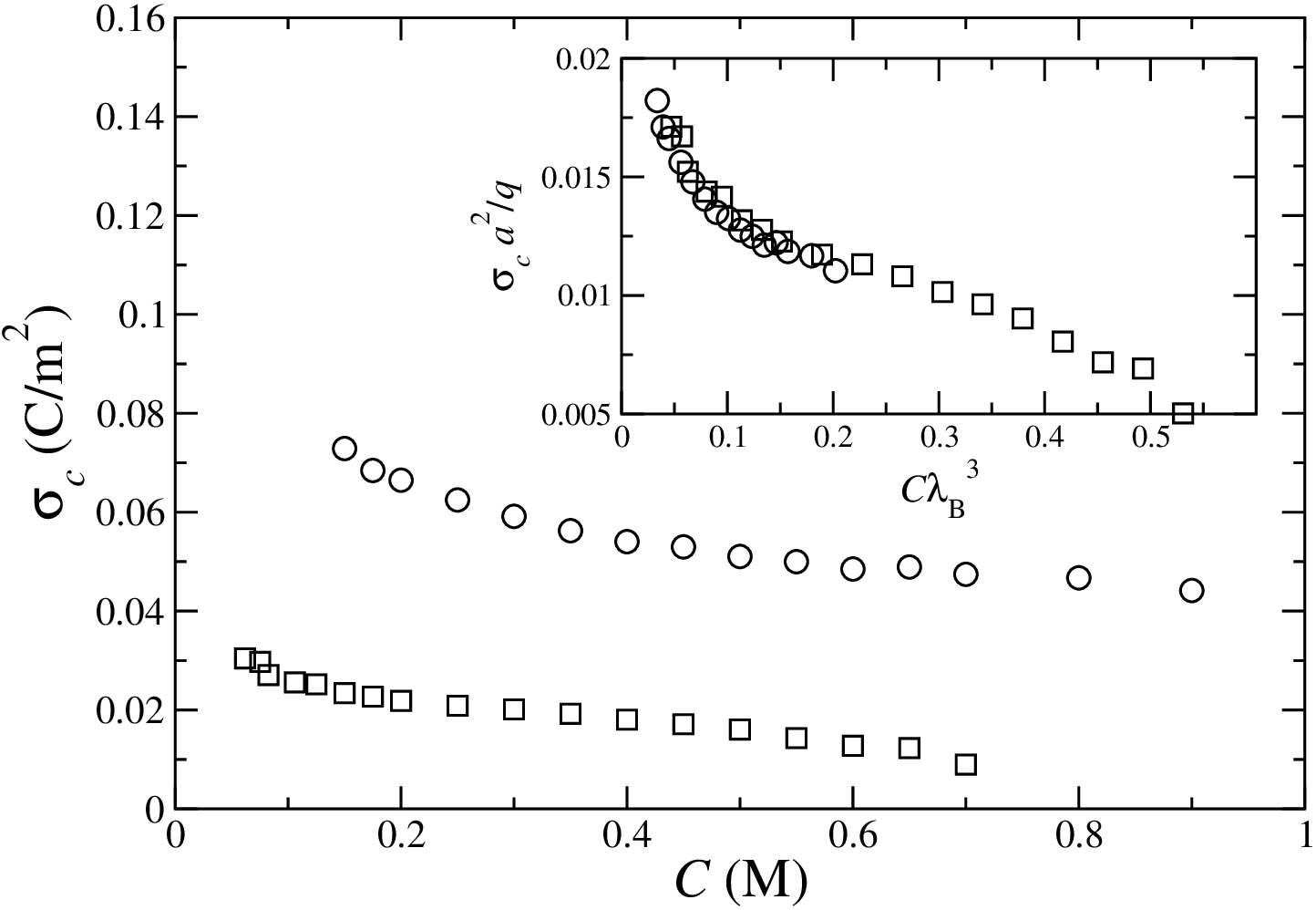}
\caption{Critical colloidal surface charge density as a function of the molar 
concentration of 3:1 electrolyte. Circles are for electrolyte with 
$a=2\,\mbox{\AA}$, $\lambda_B=7.2\,\mbox{\AA}$; squares are for
electrolyte with $a=3\,\mbox{\AA}$, $\lambda_B=10.8\,\mbox{\AA}$. 
In both cases the ratio $a/\lambda_B = 0.278$ is the same. 
Inset shows the data collapse when
the concentrations and $\sigma_c$ are properly scaled.}
\label{fig2}
\end{figure}

We next studied the dependence of the critical surface charge density on the concentration of
electrolyte.  In Fig. 2, $\sigma_c$ is plotted as a function of $C$ for two 
different electrolytes: $a=2\,\mbox{\AA}$, $\lambda_B=7.2\,\mbox{\AA}$; and  
$a=3\,\mbox{\AA}$, $\lambda_B=10.8\,\mbox{\AA}$. Although clearly distinct, the two systems
are ``similar'', since the ratio $\lambda_B/a$ is the same in both cases. Therefore, if
$\sigma_c a^2/q$ is plotted as a function of $\lambda_B C^{1/3}$ 
( or as a function of $C \lambda_B^3$) the data for the two systems
should collapse onto a single curve.  This is precisely what is found, see the inset of Fig. 2.\\  

\begin{figure}[t]
\includegraphics[width=8cm]{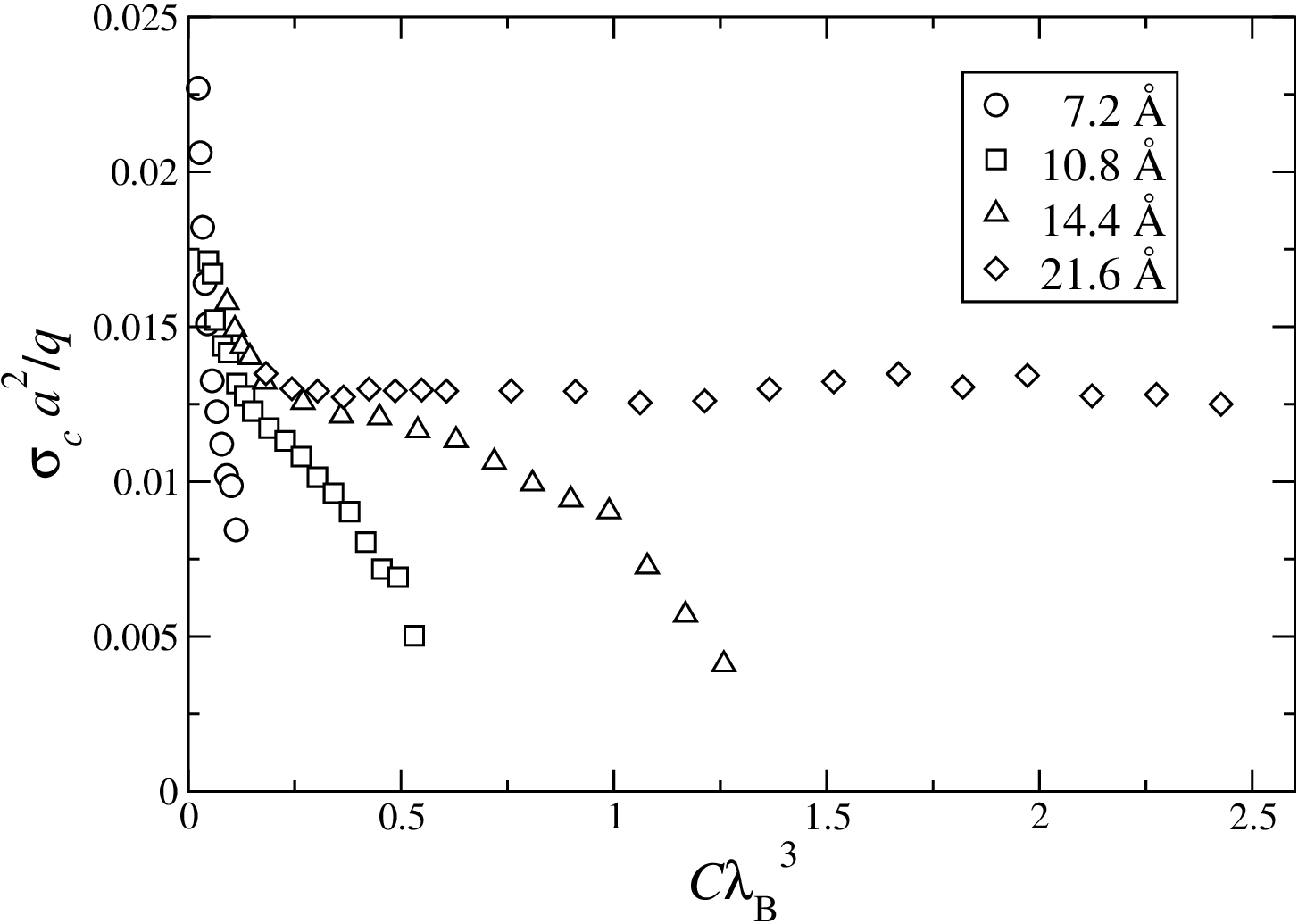}
\caption{Scaled critical colloidal surface charge density as a function of the scaled 
concentration of 3:1 electrolyte. The microions are of radius $a=3\,$\AA.}
\label{fig3}
\end{figure}

If the electrolyte systems are not connected by the similarity transformation, a data
collapse is not expected. Nevertheless, as was discussed in the previous section, if $a/\lambda_B$ is
small, $g(x\approx0,y)$ should be nearly constant.  We expect, however, 
the deviation from this constancy
to take place when the product $xy$ becomes sufficiently large --- 
the separation between the microions becomes compatible to the ionic size, or when the concentration
becomes so small that the entropic effects dominate over the electrostatics. In Fig. 3
we plot $\sigma_c a^2/q$ as a function $C \lambda_B^3$, for various electrolytes of different 
$\lambda_B$.\\

\vspace*{0.5cm}
\begin{figure}[th]
\includegraphics[width=8cm]{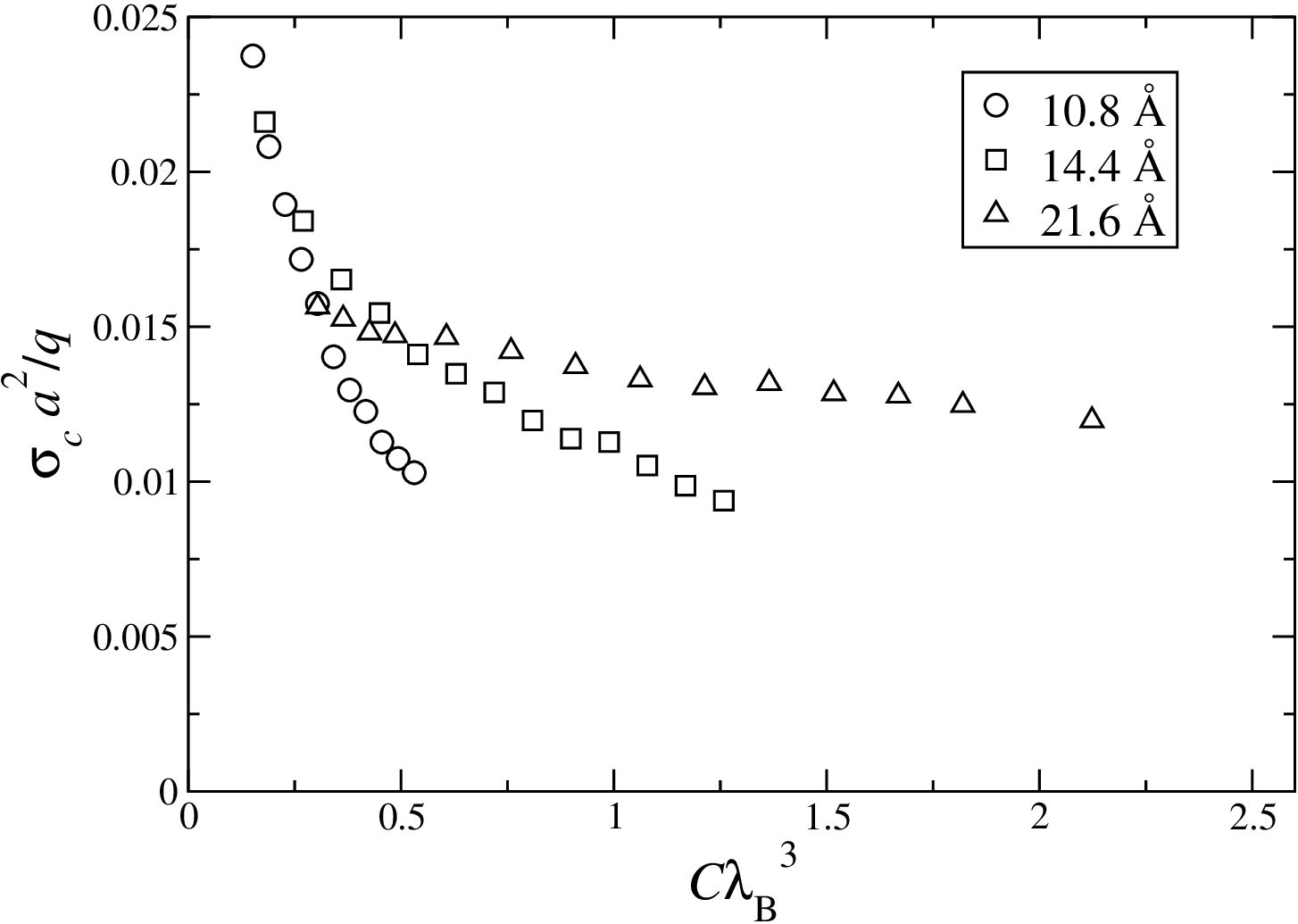}
\caption{Scaled critical colloidal surface charge density as a function of the scaled
concentration of 2:1 electrolyte. The microions have radius $a=2\,$\AA.}
\label{fig4}
\end{figure}

The figure clearly shows the inflection point of $\sigma_c$ as a function of the electrolyte
concentration.  Furthermore, in the limit of 
vanishing $a/\lambda_B$, the inflection point turns into 
plateau which extends up to the concentrations for which $C^{1/3} a \approx 0.2$, consistent with the
discussion presented earlier.  In the limit of vanishing ionic size, the plateau region extends 
indefinitely.\\ 

\section{Conclusions}
\label{conclusions}
Using extensive Monte Carlo simulations we have 
studied the dependence of the minimal colloidal charge at which the reversal of 
the electrophoretic mobility first
takes place, on the concentration of 3:1 electrolyte and on the ionic size.  
The critical surface charge density $\sigma_c$ was found to exhibit 
an inflection point as a function of the 
electrolyte concentration.  In the limit of small  $a/\lambda_B$, the inflection point  
becomes a flat plateau, extending from the  lower concentration
$C_l$ to the upper concentration $C_u$.  The value of the lower bound is delimited by the distances
at which the entropic effects begin to
dominate over the electrostatics, $C_l^{1/3} \lambda_B \approx 0.5$,
while the value of the upper bound is determined by the distances at which the hard-core repulsion 
begins to dominate over everything $C_u^{1/3} a \approx 0.2$.  
In the interval $[C_l,C_u]$ the surface charge density
is found to be $\sigma_c \approx 0.013 q/a^2$, independent of the electrolyte concentration. The 
plateau disappears, turns into a simple inflection point with vanishing second derivative, 
when $ \lambda_B/a<2.5$.  The same behavior was found to occur 
for electrolytes of other valences.  For 
example, in Fig. 4 we plot the critical surface charge density as a function the 
electrolyte concentration for 2:1 electrolyte.  
Once again the inflection point and the 
formation of the plateau are evident.  However, in this case, 
to compensate for the weaker electrostatic interactions between
the coions and the counterions,
the value of $\lambda_B$ must be significantly lower for the plateau to appear clearly.
What is surprising, however,  is that the scaled surface charge
density $\sigma_c a^2/q$, appears {\bf not} to depend on the valence of the 
electrolyte --- or depend only very weakly --- 
in the plateau region.  In the case of 2:1 electrolyte we find that    
$\sigma_c a^2/q \approx 0.015$ as compared to the $0.013$ for the 3:1 electrolyte.  It will be
interesting to see if this curious behavior 
persists for other values of $z$.  At the moment, there is
no theory which can quantitatively account for these curious findings. 
We hope that the present study will provide a simulational 
benchmark against which the future theoretical
predictions can be tested.   

\section{Acknowledgments}
This work is partially supported by CNPq and by the
US-AFOSR under the grant FA9550-06-1-0345.

\end{document}